
\input harvmac.tex
\noblackbox
\def\boxit#1{\vbox{\hrule\hbox{
\vrule\kern3pt\vbox{\kern3pt#1\kern3pt}\kern3pt\vrule}\hrule}}
%
\Title{\vbox{\baselineskip12pt
\hbox{NI-94-012}
\hbox{DAMTP/R 94-26}
\hbox{UCSBTH-94-25}
\hbox{gr-qc/9409013}}}
{\vbox{\centerline{Entropy, Area, and Black Hole Pairs}}}

\baselineskip=12pt
\centerline{S. W. Hawking,$^{1}$\footnote {$^*$}{Current address:
Department of Applied Mathematics and Theoretical Physics, Silver St.,
Cambridge CB3 9EW. Internet: swh1@amtp.cam.ac.uk}
Gary T. Horowitz,$^{1}$\footnote {$^{\dagger}$} {Current address:
Physics Department, University of California, Santa Barbara, CA. 93111.
Internet: gary@cosmic.physics.ucsb.edu} and Simon F. Ross$^2$}
\bigskip
\centerline{\sl $^1$ Isaac Newton Institute for Mathematical Sciences}
\centerline{\sl University of Cambridge, 20 Clarkson Rd., Cambridge CB3 0EH}

\bigskip
\centerline{\sl $^2$Department of Applied Mathematics and Theoretical Physics}
\centerline{\sl University of Cambridge, Silver St., Cambridge CB3 9EW}
\centerline{\sl Internet: S.F.Ross@amtp.cam.ac.uk}

\bigskip
\centerline{\bf Abstract}
\medskip

We clarify the relation between gravitational entropy and the area of
horizons.
We first show that the entropy of an extreme Reissner-Nordstr\"om
black hole is $zero$, despite the fact that its horizon has nonzero area.
Next, we
consider the pair creation of extremal and nonextremal black holes.
It is shown that the action which governs the rate of this pair creation
is directly related to the area of the acceleration horizon and (in
the nonextremal case) the area of
the black hole event horizon. This provides a simple explanation of the
result that the rate of pair creation of non-extreme black holes is
enhanced by precisely the black hole entropy.
Finally, we discuss black hole $annihilation$,
and argue that Planck scale remnants are not sufficient to preserve unitarity
in quantum gravity.

\Date{September 1994}
\def\b{\beta}

\def\p{\partial}
\def\d{\nabla}
\def\A{{\cal A}}
\def\H{{\cal H}}

\def\S{\Sigma}
\def\RN{Reissner-Nordstr\"om}
\def\vp{\varphi}
\def\({\left (}
\def\){\right )}
\def\[{\left [}
\def\]{\right ]}

\def\np {  Nucl. Phys. }
\def \pl { Phys. Lett. }

\def \prl { Phys. Rev. Lett. }
\def \pr  { Phys. Rev. }

\def \cmp {Commun. Math. Phys.}
\gdef \jnl#1, #2, #3, 1#4#5#6{ {\sl #1~}{\bf #2} (1#4#5#6) #3}
\lref\haho{ S.W. Hawking and G. Horowitz, ``Gravitational Hamiltonians,
Actions, and Surface Terms", to appear.}
\lref\hawking{ S. W. Hawking, \jnl \cmp, 43, 199, 1975.}
\lref\hawk{S.W. Hawking, in {\it General Relativity, an Einstein Centenary
Survey}, eds. S. W. Hawking and W. Israel (Cambridge University Press) 1979.}
\lref\ggs{D. Garfinkle, S.B. Giddings and A. Strominger, \jnl \pr,
D49, 958, 1994.}
\lref\garstrom{D. Garfinkle and A. Strominger, \jnl \pl, B256, 146, 1991.}
\lref\dgkt{H.F. Dowker, J.P. Gauntlett, D.A. Kastor and J. Traschen,
\jnl \pr,  D49, 2909,  1994.}
\lref\dggh{ H.F. Dowker, J.P. Gauntlett, S.B. Giddings and G.T. Horowitz,
\jnl \pr, D50, 2662, 1994.}
\lref\ernst{F. J. Ernst, \jnl J. Math. Phys., 17, 515, 1976.}
\lref\schwinger{J. Schwinger, \jnl \pr, 82, 664, 1951.}
\lref\gwg{G.W. Gibbons,
in {\sl Fields and geometry}, proceedings of
22nd Karpacz Winter School of Theoretical Physics: Fields and
Geometry, Karpacz, Poland, Feb 17 - Mar 1, 1986, ed. A. Jadczyk (World
Scientific, 1986).}
\lref\gm{G.W. Gibbons and K. Maeda,
\jnl \np,  B298, 741, 1988.}
\lref\melvin{M. A. Melvin, \jnl \pl, 8, 65, 1964.}
\lref\afma{I.K. Affleck and N.S. Manton, \jnl \np, B194, 38, 1982;
I.K. Affleck, O. Alvarez, and N.S. Manton, \jnl \np, B197, 509, 1982.}
\lref\wald{ R. Wald, {\it General Relativity}, University of Chicago
Press, 1984.}
\lref\teit{M. Banados, C. Teitelboim, and J. Zanelli, \jnl \prl, 72, 957,
1994.}

\lref\bch{ J. Bardeen, B. Carter, and S. Hawking, \jnl \cmp, 31, 161, 1973;
J. Bekenstein, \jnl \pr, D7, 2333, 1973.}
\lref\trivedi{S. Trivedi, \jnl \pr, D47, 4233, 1993.}
\lref\gk{G.  W.  Gibbons and R.  Kallosh, paper in preparation}

\lref\by{J.D. Brown, E.A. Martinez and J.W. York, \jnl \prl, 66, 2281, 1991;
J.D. Brown and J.W. York, {\sl Phys. Rev.} {\bf D47} (1993) 1407, 1420.}
\newsec{Introduction}

The discovery of black hole radiation \hawking\
confirmed earlier indications \bch\ of
a close link between
thermodynamics and black hole physics. Various arguments were given that a
black hole has an entropy which is  one quarter of the area of its
event horizon in Planck units. However,  despite extensive discussion, a
proper understanding of this entropy is still lacking. In particular
there is no direct connection between this entropy and the
`number of internal states' of a black hole.

We will re-examine the connection between gravitational entropy and horizon
area
in two different contexts.
We first consider charged black holes and show that while non-extreme
configurations satisfy the usual relation $S=\A_{bh}/4$, extreme \RN\
black holes do not. They always have zero entropy even though
their event horizon has nonzero area.
The entropy changes discontinuously
when the  extremal limit is reached. We will see that this is a result
of the fact that the horizon is infinitely far away for extremal holes
which results in a change in the topology of the Euclidean solution.

The second context is quantum pair creation
of black holes.
It has been known for some time that one can create pairs of oppositely charged
GUT monopoles
in a strong background magnetic field \afma.
The rate for this
process can be calculated in an instanton approximation and is given
by $e^{-I}$ where $I$ is the Euclidean action of the instanton. For monopoles
with mass $m$ and charge $q$, in a background field $B$ one finds (to leading
order in $qB$) that
$I=\pi m^2/qB$. It has recently
been argued that charged black holes can similarly be pair created in
a strong  magnetic field \refs{\gwg,\garstrom,\dggh}.
 An appropriate instanton has been found and
its action computed.
The instanton is
obtained by starting with a solution to the Einstein-Maxwell equations
found by Ernst \ernst, which describes oppositely charged black holes
uniformly accelerated in a background magnetic field. This solution has
a boost symmetry which becomes null on an acceleration horizon as well
as the black hole event horizon, but is time-like in between. One can
thus analytically continue to obtain the Euclidean instanton.
It turns out that regularity of the instanton requires that
the black holes are either extremal or slightly nonextremal.
In the nonextremal case,
 the two black hole event horizons are
identified to form a wormhole in space.
It was shown in \dggh\ that the action for the instanton creating
extremal black holes
is identical to that creating gravitating monopoles \ggs\ (for small $qB$)
while the
action for non-extreme black holes
is less  by precisely the entropy of one black
hole ${\cal A}_{bh}/4$. This implies that
 the pair creation
rate for non-extremal black holes is enhanced over that of
extremal black holes by a factor of $e^{{\cal A}_{bh}/4}$, which may be
interpreted as saying that non-extreme black holes have
$e^{{\cal A}_{bh}/4}$ internal states and are produced in correlated pairs,
while the extreme
black holes have a unique internal state. This was not understood at the time,
but is in perfect agreement with our result that the entropy of
extreme black holes is zero.

To better understand the rate of pair creation, we relate the instanton
action to an energy associated with boosts, and surface terms at the horizons.
While the usual energy is unchanged in the pair creation process, the boost
energy need not be. In fact, we will see that it is changed in the
pair creation of nongravitating GUT monopoles. Remarkably, it turns out
that it is unchanged when gravity is included. This allows us to derive
a simple formula for the instanton action. For the pair creation of
nonextremal black holes we find
\eqn\keyeq{ I= -{1\over 4} (\Delta \A + \A_{bh}), }
where $\Delta \A$ is the difference between the area of the acceleration
horizon
when the black holes are present and when they are absent,
and $\A_{bh}$ is the area of the black hole horizon.
For the pair creation of extremal black holes (or gravitating monopoles)
the second term is absent so the rate is entirely determined by the area
of the acceleration horizon,
\eqn\keyeqq{ I= -{1\over 4} \Delta \A. }
This clearly shows the origin of the fact that nonextremal black holes
are pair created at a higher rate given by the entropy of one
black hole.

The calculation of each side of \keyeqq\ is rather subtle. The area of
the acceleration horizon is infinite in both the background magnetic field
and the Ernst solution. To compute the finite change in area we first
compute the
area in the Ernst solution out to a large circle. We then subtract off  the
area in the background
magnetic field solution out to a circle which is chosen to have
the same proper length and the same value of $\oint A$ (where $A$ is the vector
potential). Similarly, the
instanton action is finite only after we subtract off the
infinite contribution coming from the background
magnetic field. In \garstrom, the calculation was done by computing the
finite change in the action when the
black hole charge is varied,  and then integrating  from zero charge to
the desired $q$. In \dggh, the action was calculated inside a large
sphere and the background contribution was subtracted using a
coordinate matching condition. Both methods yield the same result.
But given the importance of the action for the pair creation rate,
one would like to have a direct calculation of it by matching the
intrinsic geometry on a boundary near infinity as has been done for other
black hole instantons.
We will present such a derivation here and show that
the result is in agreement with
the earlier approaches. Combining this with our calculation of $\Delta \A$,
we explicitly confirm the relations \keyeq\ and \keyeqq.

Perhaps the most important application of gravitational entropy
is to the `black hole information puzzle'.
Following the discovery of black hole radiation, it was argued that
information and quantum coherence can be lost in quantum gravity.
This seemed to be an inevitable consequence of the semiclassical
calculations which showed that black holes emit thermal radiation
and slowly evaporate.  However, many people find it difficult to accept
the idea of nonunitary evolution. They have suggested that either the
information thrown into a black hole comes out in detailed correlations
not seen in the semiclassical approximation, or that the endpoint of
the evaporation is a Planck scale remnant which stores the missing
information. In the latter case, the curvature outside the horizon
would be so large that semiclassical arguments would no longer be valid.
However, consideration of black hole pair creation suggests
another quantum gravitational process involving
black holes, in which information seems to be lost yet the curvature outside
the horizons always remains small.

The basic observation is that
if black holes can be pair created, then it must be possible for them
to annihilate.
In fact, the same instanton which describes black hole
pair creation can also be interpreted as describing black hole annihilation.
Once one accepts the idea that black holes can  annihilate, one can
construct an argument for information loss as follows. Imagine pair
creating two magnetically charged (nonextremal)
black holes which move far apart
into regions of space without a  background
magnetic field. One could then treat each black hole
independently and  throw an
arbitrarily large amount of matter and information into them. The holes
would then radiate and return to their original mass.
One could then bring the two holes back
together again and try to annihilate them. Of course, there is always the
possibility that
they will collide and form a
black hole with no magnetic charge and  about double the horizon area. This
black hole could evaporate in the usual way down to Planck scale curvatures.
However,
there is a probability of about $e^{-\A_{bh}/4}$
times the monopole annihilation probability that the black holes will simply
annihilate, their energy being given off
as electromagnetic or gravitational radiation.
One can choose the magnetic field and the value of the magnetic charge
in such a way that the curvature is everywhere small. Thus the semiclassical
approximation should remain valid.
This implies that even if small black hole
remnants exist, they are not sufficient to preserve unitarity.
This discussion applies to nonextremal black holes.
Since extremal black holes have zero entropy, they
behave differently, as we will explain.

In the next section  we discuss the entropy of a single static black hole
and show
that an extreme \RN\ black hole has zero entropy. Section 3 contains
a review of the Ernst instanton which describes pair creation of extremal
and nonextremal black holes. In section 4 we discuss the boost energy
and show that while it is changed for pair creation in flat space, it
is unchanged for pair creation in general relativity. Section 5 contains a
derivation of the relations \keyeq\ and \keyeqq\ and the detailed calculations
of the acceleration horizon area and instanton action which confirm them.
Finally, section 6 contains further discussion of black hole annihilation
and some concluding remarks.

In Appendix A we consider the
 generalization of the Ernst instanton which includes an arbitrary coupling
to a dilaton \dgkt. We will extend the development of the preceding
sections to this case, showing that the boost energy is still
unchanged in this case, and calculating the difference in area and the
instanton action using appropriate boundary conditions.
The result for the instanton action is in complete agreement with \dggh.


\newsec{Extreme Black Holes Have Zero Entropy}

In this section we consider the entropy of a single static black hole.
The reason that gravitational
configurations can have nonzero entropy is that the Euclidean solutions
can have nontrivial topology \hawk. In other words, if we start with a static
spacetime and identify imaginary time with period $\b$, the manifold
need not have topology $S^1 \times \S$ where $\S$ is some three manifold.
In fact, for non-extreme black holes, the topology is
$S^2 \times R^2$. This means that the foliation one introduces to
rewrite the action in Hamiltonian form must meet at a two sphere $S_h$.
The Euclidean Einstein-Maxwell action includes a surface
term,
\eqn\eaction{ I = {1\over 16 \pi} \int_M (-R + F^2) -{1\over 8\pi}\oint_{\p M}
   K,}
where $R$ is the scalar curvature, $F_{\mu\nu}$ is the Maxwell field,
and $K$ is the trace of the extrinsic curvature of the boundary.
In fact, if the spacetime is noncompact, the action is defined only relative to
some background solution $(g_0,F_0)$.
 This background is usually taken to be flat space with zero field, but we
shall consider more general asymptotic behavior.
When one rewrites the action in Hamiltonian form, there is an extra
 contribution from the two sphere
 $S_h$. This arises since the surfaces of constant time meet at $S_h$ and the
 resulting corner gives a  delta-function contribution to $K$ \hawk.
 Alternatively, one can calculate the contribution from $S_h$ as follows \teit\
  (see \by\ for another approach).
 The total action can be written as
the sum of the action of a small tubular neighborhood of $S_h $ and everything
outside. The action for the region outside
reduces to the standard Hamiltonian form\foot{The surface terms in the
Hamiltonian can be obtained directly from the surface terms in the action.
For a detailed discussion which includes spacetimes which are not
asymptotically flat (e.g. the Ernst solution) and horizons which are not
compact (e.g. acceleration horizons) see \haho.}, which for a static
configuration
yields the familiar result $\b H $. The action for the small neighborhood
of $S_h $ yields $-\A_{bh}/4$
where $\A_{bh}$ is
the area of $S_h$.
Thus the total Euclidean action is
\eqn\impeq{ I = \b H -{1\over 4} \A_{bh}.}

The usual thermodynamic formula for the entropy is
\eqn\othermo{ S = - \left( \beta {\partial \over \partial \beta} - 1\right)
\log Z, }
where the partition function $Z$ is given (formally) by the integral
of $e^{-I}$ over all Euclidean configurations which are periodic in
imaginary time with period $\b$ at infinity. The action for the
solution describing a nonextremal black hole is \impeq\ so if we
approximate $\log Z \approx -I$, we obtain the usual result
\eqn\entropy{ S = {1\over 4} \A_{bh}. }

Recall that the \RN\ metric is given by
\eqn\exrn{ ds^2 = -\( 1-{2M\over r} + {Q^2 \over r^2}\) dt^2 +
     \( 1-{2M\over r} + {Q^2 \over r^2}\)^{-1} dr^2 +r^2 d\Omega.}
For non-extreme black holes $Q^2 < M^2$, the above discussion
applies. But the extreme \RN\ solution is qualitatively
different. When $Q^2 = M^2$, the horizon $r=M$ is infinitely far away
along spacelike directions.  In the Euclidean solution, the horizon is
infinitely far away along all directions.  This means that the
Euclidean solution can be identified with any period $\b$. So the
action must be proportional to the period $I \propto \beta$. It
follows from \othermo\ that in the usual approximation $\log Z \approx
-I$, the entropy is zero,
\eqn\extre{  S_{extreme} = 0.}
This is consistent with the fact that gravitational entropy should be
associated with nontrivial topology. The Euclidean extreme \RN\
solution (with $\tau$ periodically identified) is topologically $S^1
\times R \times S^2$.  Since there is an $S^1$ factor, the surfaces
introduced to rewrite the action in canonical form do not
intersect. Thus there is no extra contribution from the horizon and the
entropy is zero. Since the area of the event horizon of an extreme
\RN\ black hole is nonzero, we conclude that {\it the entropy of a
black hole is not always equal to} $\A_{bh}/4$; \entropy\ holds only for
nonextremal black holes.

The fact that the entropy changes discontinuously in the extremal
limit implies that one should regard non-extreme and extreme black
holes as qualitatively different objects.  One is already used to the
idea that a non-extreme black hole cannot turn into an extreme hole:
the nearer the mass gets to the charge the lower the temperature and
so the lower the rate of radiation of mass. Thus the mass will never
exactly equal the charge. However, the idea that extreme and
non-extreme black holes are distinct presumably also implies that
extreme black holes cannot become non-extreme. At first sight this
seems contrary to common sense.  If one throws matter or radiation
into an extreme black hole, one would expect to increase the mass and
so make the hole non-extreme. However, the fact that one can identify
extreme black holes with any period implies that extreme black holes
can be in equilibrium with thermal radiation at any temperature.  Thus
they must be able to radiate at any rate, unlike non-extreme black
holes, which can radiate only at the rate corresponding to their
temperature. It would therefore be consistent to suppose that extreme
black holes always radiate in such a way as to keep themselves extreme
when matter or radiation is sent into them.

{}From all this it might seem that extreme and nearly extreme black
holes would appear very different to outside observers. But this need
not be the case. If one throws matter or radiation into a nearly
extreme black hole, one will eventually get all the energy back in
thermal radiation and the hole will return to its original
state. Admittedly, it will take a very long time, but there is no
canonical relationship between the advanced and retarded time
coordinates in a black hole. This means that if one sends energy into
an extreme black hole there is no obviously preferred time at which
one might expect it back. It might therefore take as long as the
radiation from nearly extreme black holes. If this were the case, a
space-like surface would intersect either the infalling matter or the
outgoing radiation just outside the horizon of an extreme black
hole. This would make its mass seem greater than its charge and so an
outside observer would think it was non-extreme.

If extreme black holes behave just like nearly extreme ones is there any way
in which we can distinguish them? A possible way would be in black hole
annihilation, which will be discussed in section 6.

Two dimensional calculations \trivedi\ have indicated that the
expectation value of the energy momentum tensor tends to blow up on
the horizon of an extreme black hole. However, this may not be the
case in a supersymmetric theory.  Thus it may be possible to have
extreme black holes only in supergravity theories in which the
fermionic and bosonic energy momentum tensors can cancel each
other. Because they have no entropy such supersymmetric black holes
might be the particles of a dual theory of gravity.

There is a problem in calculating the pair creation of extreme black
holes even in supergravity.  As Gibbons and Kallosh \gk\ have pointed
out, one would expect cancellation between the fermionic and bosonic
energy momentum tensors only if the fermions are identified
periodically.  In the Ernst solution however, the presence of the
acceleration horizon means that the fermions have to be antiperiodic.
Thus it may be that the pair creation of extreme black holes will be
modified by strong quantum effects near the horizon.


\newsec{The Ernst Solution}


The solution describing a background magnetic field in general
relativity is  Melvin's magnetic universe \melvin,
\eqn\melvins{
\eqalign{
&ds^2=\Lambda^2 \left[-dt^2+dz^2+d\rho^2\right]
+\Lambda^{-2}\rho^2 d\varphi^2, \cr
&A_\varphi={\widehat{B}_M\rho^2\over 2\Lambda}, \qquad \Lambda=1+{1\over
4}\widehat{B}_M^2\rho^2\ .\cr} }
The Maxwell field is $F^2 = 2\widehat{B}_M^2/\Lambda^4$, which is a maximum on
the
axis $\rho=0$ and decreases to zero at infinity. The parameter $\widehat{B}_M$
gives the value  of the magnetic field on the axis.

The Ernst solution is given by
\eqn\ernsts{
\eqalign{
&ds^2=(x-y)^{-2}A^{-2}\Lambda^2
\left[G(y)dt^2-G^{-1}(y)dy^2
+G^{-1}(x)dx^2\right] \cr &\qquad +
(x-y)^{-2}A^{-2}\Lambda^{-2}G(x) d\varphi^2, \cr
&A_\varphi=-{2\over B\Lambda}\[1+{1\over 2}Bqx\]
+k, }}
where the functions $\Lambda \equiv\Lambda(x,y)$,
  and $G(\xi)$ are
\eqn\fns{\eqalign{
&\Lambda=\[1+{1\over 2}Bqx\]^2+{B^2\over 4A^2(x-y)^2}G(x), \cr
&G(\xi)=(1+r_-A\xi)(1-\xi^2-r_+A\xi^3),\cr}}
and $q^2 = r_+ r_-$. This solution represents two oppositely charged
black holes uniformly accelerating in a background magnetic field.

It is convenient to set $\xi_1 = -1/(r_- A) $ and let
$\xi_2\le\xi_3<\xi_4$ be the three roots of the cubic factor in $G$. The
function $G(\xi)$ may then be written as
\eqn\groots{
G(\xi)=-(r_+A)(r_-A)
(\xi-\xi_1)(\xi-\xi_2)(\xi-\xi_3)(\xi-\xi_4) .}
We  restrict $\xi_3 \leq x \leq \xi_4$ in order for
the metric to have Lorentz signature. Because of the conformal factor
$(x-y)^{-2}$ in the metric, spatial infinity is reached when $x,y
\rightarrow \xi_3$, while $y \rightarrow x$ for $x \neq \xi_3$
corresponds to null or time-like infinity. The range of $y$ is therefore
$-\infty < y < x$. The axis $x=\xi_3$ points towards spatial infinity,
and the axis $x=\xi_4$ points towards the other black hole. The
surface $y=\xi_1$ is the inner black hole horizon, $y = \xi_2$ is the
black hole event horizon, and $y=\xi_3$ the acceleration horizon. We
can choose $\xi_1 < \xi_2$, in which case the black holes are
non-extreme, or $\xi_1 = \xi_2$, in which case the black holes are
extreme.

As discussed in \dgkt, to ensure that the metric is free of conical
singularities at both poles, $x=\xi_3, \xi_4$, we must impose the
condition
\eqn\nonodes{G^\prime(\xi_3)\Lambda(\xi_4)^2
= -
G^\prime(\xi_4)\Lambda(\xi_3)^2,}
where $\Lambda(\xi_i)\equiv \Lambda(x=\xi_i)$. For later convenience, we define
$L \equiv \Lambda (x=\xi_3)$. When \nonodes\ is
satisfied, the spheres are regular as long as $\vp$ has period
\eqn\phiperiod{\Delta\vp={4\pi L^2\over
G^\prime(\xi_3) }\ .}

We choose the
constant $k$ in \ernsts\ to be $k = 2/ B L^{1/2}$ so as to confine the
Dirac string of the
magnetic field to the axis $x= \xi_4$.  We define a physical magnetic
field parameter $\widehat{B}_E = B G'(\xi_3) / 2 L^{3/2}$, which
is the value of the magnetic field on the axis at infinity. The physical
charge of the black hole is defined by
\eqn\pcharge{\widehat{q} = {1 \over 4\pi} \int F =  q
{ L^{3 \over 2} (\xi_4-\xi_3) \over G'(\xi_3) (1+ {1 \over 2}
qB \xi_4)}\ .}
If we also define $m = (r_+ + r_-)/2$, we can see that the solution
\ernsts\ depends on four parameters: the physical magnetic field
$\widehat{B}_E$, the physical magnetic charge $\widehat q$, and $A$
and $m$, which may be loosely interpreted as measures of the
acceleration and the mass of the black hole.

If we set the black hole parameters $m$ and $q$ (or equivalently, $r_+, r_-$)
to zero in \ernsts\ we obtain
\eqn\accmel{\eqalign{ds^2 = &{\Lambda^2
\over A^2 (x-y)^2} \left[ (1-y^2)
 dt^2 - {dy^2\over  (1-y^2)} + {dx^2\over
(1-x^2) }\right]\cr & + {1-x^2 \over \Lambda^2 A^2 (x-y)^2}
d\vp^2,}}
with
\eqn\Llim{\Lambda = 1+ {\widehat{B}_E^2\over 4} {1-x^2\over A^2
(x-y)^2}\ .}
This is just the Melvin metric \melvins\ expressed in
accelerated coordinates, as can be seen by the coordinate
transformations \dggh
\eqn\coordlim{\rho^2 = {1-x^2 \over  (x-y)^2 A^2}\ ,\ \eta^2 = {y^2-1\over
(x-y)^2 A^2}.}
Note that now the acceleration parameter $A$ is no longer
physical, but represents a choice of coordinates. The gauge field also
reduces to the Melvin form $A_\varphi = \widehat{B}_E \rho^2 /
2\Lambda$.
One can show \refs{\dgkt,\dggh} that the Ernst solution
reduces to the Melvin solution at large spatial distances, that
is, as $x,y \rightarrow \xi_3$.

We now turn to the consideration of the Euclidean section of the Ernst
solution, which will form the instanton. We Euclideanize \ernsts\ by
setting $\tau = it$.  In the non-extremal case, $\xi_1 < \xi_2$, the
range of $y$ is taken to be $\xi_2 \leq y \leq \xi_3$ to obtain a
positive definite metric (we assume $\xi_2 \ne \xi_3$).
To avoid conical singularities at the
acceleration and black hole horizons, we take the period of $\tau$ to
be
\eqn\pert{\beta = \Delta \tau = {4 \pi \over G'(\xi_3)}}
and require
\eqn\nost{G'(\xi_2)  = -G'(\xi_3),}
which gives
\eqn\nostrut{ \left( \xi_2-\xi_1
\over \xi_3 -\xi_1 \right) (\xi_4-\xi_2) = (\xi_4-\xi_3)
.}
This condition can be simplified to
\eqn\nostrtt{ \xi_2 - \xi_1 = \xi_4 - \xi_3 .}
The resulting
instanton has topology $S^2 \times S^2 -\{pt\}$, where the point
removed is $x=y=\xi_3$. This instanton is interpreted
as representing the pair creation of two oppositely charged black
holes connected by a wormhole.

If the black holes are extremal, $\xi_1=\xi_2$, the black hole event
horizon lies at infinite spatial distance from the acceleration
horizon, and gives no restriction on the period of $\tau$. The range
of $y$ is then $\xi_2 < y \leq \xi_3$, and the period of $\tau$ is
taken to be \pert. The topology of this instanton  is $R^2 \times S^2
- \{ pt \}$, where the removed point is again $x=y=\xi_3$. This instanton is
interpreted as representing the pair creation of two extremal black
holes with infinitely long throats.

\newsec{Boost Energy}

Consider the pair creation of (non-gravitating) GUT monopoles in flat
spacetime. In this process the usual energy is unchanged. If the
background magnetic field extends to infinity, this energy will, of
course, be infinite. But even if it is cut off at a large distance,
the energy is conserved since in the Euclidean solution, $\d^\mu
(T_{\mu\nu} t^\nu)=0$, where $T_{\mu\nu}$ is the energy momentum
tensor and $t^\nu$ is a time translation Killing vector.  Thus the
initial energy, which is the integral of $T_{\mu\nu} t^\mu t^\nu$ over
a surface in the distant past, must equal the energy after the
monopoles are created.  However, now consider the energy associated
with a boost Killing vector in the Lorentzian solution. This
corresponds to a rotation $\xi^\mu$ in the Euclidean instanton. So the
associated energy is
\eqn\bengy{ E_B = \int_\Sigma T_{\mu\nu} \xi^\mu d\Sigma^\nu,}
where the integral is over a surface $\S$ which starts at the
acceleration horizon where $\xi^\mu = 0$ and extends to infinity.
While the vector $T_{\mu\nu} \xi^\mu$ is still conserved, which
implies that $E_B$ is unchanged under continuous deformations of $\S$
that preserve the boundary conditions, this is not sufficient to prove
that $E_B$ is unchanged in the pair creation process. This is because
every surface which starts at the acceleration horizon in the
instanton always intersects the monopole, and cannot be deformed into
a surface lying entirely in the background magnetic field.  In fact,
it is easy to show that $E_B$ {\it is} changed.   Since the analytic
continuation of the boost parameter is periodic with period $2\pi$,
the Euclidean action is just $ I = 2\pi E_B$. So the fact that the
instanton describing the pair creation of monopoles has a different
action from the uniform magnetic field means that the boost energy is
different.

We now turn to the case of pair creation of gravitating monopoles, or
black holes. The gravitational Hamiltonian is only defined with respect
to a background spacetime, and can be expressed  \haho\ (this form
of the surface term at infinity is also discussed in \by)
\eqn\hamil{ H = \int_{\Sigma}
   N\H - {1\over 8\pi} \int_{S^\infty} N({}^2 K - {}^2K_0),}
where $N$ is the lapse, $\H$ is the Hamiltonian constraint, ${}^2 K$
is the trace of the two dimensional extrinsic curvature of the
boundary near infinity, and ${}^2K_0$ is the analogous quantity for
the background spacetime.  Since the volume term is proportional to
the constraint, which vanishes, the energy is just given by a surface
term at infinity. The Hamiltonian for Melvin is zero since we are
using it as the background in which $^2 K_0$ is evaluated.  We now
calculate the Hamiltonian for the Ernst solution and show that it is
also zero. Thus the boost energy is unchanged by pair creation in the
gravitational case.

Since the spacetime is noncompact, we have to take a boundary `near
infinity', and eventually
take the limit as it tends to infinity. The surface $\Sigma$ in the
Ernst solution is a surface of constant $t$ in the Ernst metric
\ernsts, running from the acceleration horizon to a boundary at large
distance. As a general principle, we want the boundary to obey the
Killing symmetries of the metric, and in this case, we choose it to be
given by $x-y= \epsilon_E$, as in \dggh. The result in the limit as
the boundary tends to infinity should be independent of this choice.

The first part of the surface term is computed in the Ernst metric,
and the second part in the Melvin metric. We need to ensure that the
boundaries that we use in computing these two contributions are
identical; that is, we must require that the intrinsic geometry and
the Maxwell field on the boundary are the same. Because
the Ernst solution reduces to the Melvin solution at large distances,
it is possible to find coordinates in which the induced metric and
gauge field on the boundary agree explicitly.

The analogue of the surface $\Sigma$ for the Melvin solution is a
surface of constant boost time $t$ of the Melvin metric in the
accelerated form \accmel. We want to find a boundary lying in this
surface with the same intrinsic geometry as the above. We will require
that the boundary obey the Killing symmetries, but there is still a
family of possible embeddings. We assume the boundary
lies at $x-y = \epsilon_M$.  It is not clear that the results will be
independent of this assumption, but this is the simplest form the
embedding in Melvin can take, so let us proceed on this basis.

If we make coordinate transformations
\eqn\cchangei{   \varphi = { 2L^2 \over
G'(\xi_3)} \varphi',\qquad t = {2 \over G'(\xi_3)} t',}
and
\eqn\intchii{ x = \xi_3 + \epsilon_E \chi, \ \ \ y = \xi_3 + \epsilon_E
(\chi-1)}
in the Ernst metric (note that $\Delta\varphi' = 2\pi$,  $0 \leq
\chi \leq 1$, and the analytic continuation of $t'$ has period $2\pi$), then
the metric on the boundary is
\eqn\bmetricE{^{(2)}  ds^2 = {2 L^2 \over A^2 \epsilon_E G'(\xi_3)}
\left\{ - {\lambda^2 d \chi^2 \over 2\chi (\chi-1)} +
\lambda^{-2} \left[ 2 \chi + \epsilon_E \chi^2 {G''(\xi_3) \over
G'(\xi_3)} \right] d \varphi'^2 \right\},}
where
\eqn\blambdaE{ \lambda = {\widehat{B}_E^2 L^2 \over A^2 G'(\xi_3)
\epsilon_E} \chi +
{\widehat{B}_E^2 L^2 G''(\xi_3) \over 2 A^2 G'(\xi_3)^2} \chi^2 +1,}
and everything is evaluated only up to second non-trivial order in
$\epsilon_E$, as higher-order terms will not contribute to the
Hamiltonian in the limit $\epsilon_E \rightarrow 0$.

Using
\eqn\chiM{x = -1+\epsilon_M \chi,\ \ \ y = -1+\epsilon_M (\chi-1),}
the metric of the boundary in Melvin is
\eqn\bmetricM{^{(2)} ds^2 = {1 \over \bar{A}^2 \epsilon_M} \left\{ - {
\Lambda^2 d \chi^2 \over 2 \chi (\chi-1)} + \Lambda^{-2} \left[ 2 \chi -
\epsilon_M \chi^2 \right] d\varphi^2 \right\},}
where
\eqn\blambdaM{\Lambda = {\widehat{B}_M^2 \over 2 \bar{A}^2 \epsilon_M} \chi -
{\widehat{B}_M^2  \over 4 \bar{A}^2} \chi^2 +1.}
Recall that $\bar{A}$ represents a choice of coordinates in the Melvin metric.

We also want to match the magnetic fields. For the
Ernst solution, the electromagnetic field at the boundary is given by
\eqn\fluxE{F_{\chi\varphi'} = {2 A^2 G'(\xi_3) \epsilon_E \over\widehat{B}_E^3
L^2 \chi^2} \left[ 1 - {2A^2 G'(\xi_3) \epsilon_E \over
\widehat{B}_E^2 L^2 \chi} \right],}
while for Melvin it is
\eqn\fluxM{F_{\chi\varphi} = {4 \bar{A}^2 \epsilon_M \over
\widehat{B}_M^3 \chi^2} \left[ 1 - {4
\bar{A}^2 \epsilon_M \over \widehat{B}_M^2 \chi} \right].}
If we fix the remaining coordinate freedom by choosing
\eqn\abar{\bar{A}^2 = -{G'(\xi_3)^2 \over 2 L^2 G''(\xi_3)} A^2,}
and  write $\epsilon_M$ and $\widehat{B}_M$ as
\eqn\expans{\epsilon_M = -{G''(\xi_3) \over G'(\xi_3)} \epsilon_E (1+
\alpha \epsilon_E), \ \ \ \widehat{B}_M = \widehat{B}_E (1+ \beta
\epsilon_E),}
then we can easily see that the induced metrics \bmetricE\ and
\bmetricM\ and the gauge fields \fluxE\ and \fluxM\ of the boundary
may be matched by taking $\alpha=\beta=0$.

Note that, for the Ernst metric, the lapse (with respect to the time
coordinate $t'$) is
\eqn\lapE{ N = \left( 4 L^2 (1-\chi)\over A^2 \epsilon_E G'(\xi_3)
\right)^{1/2} \lambda \left[ 1 + \epsilon_E (\chi-1) {G''(\xi_3) \over
4 G'(\xi_3)} \right],}
where $\lambda$ is given by \blambdaE. For the Melvin metric, the
lapse (with respect to the boost time $t$ appearing in \accmel) is
\eqn\lapM{ N = \left( 2 (1-\chi) \over \bar{A}^2 \epsilon_M
\right)^{1/2} \Lambda \left[ 1 - {1 \over 4} \epsilon_M (\chi-1)
\right],}
where $\Lambda$ is given by \blambdaM. We therefore see that the lapse
functions are also matched by taking $\alpha=\beta=0$.

We may now calculate the extrinsic curvature $^2 K$ of the boundary
embedded in the Ernst solution, which gives
\eqn\excurv{\int_{S^\infty} N\,  {}^2 K = {8 \pi L^2 \over A^2
\epsilon_E G'(\xi_3)} \left[1- {1 \over 4} \epsilon_E {G''(\xi_3)
\over G'(\xi_3)} \right].}
Calculating the extrinsic curvature
$^2 K_0$ of the boundary embedded in the Melvin solution gives
\eqn\excurvM{\int_{S^\infty} N\,  {}^2 K_0 = {4 \pi \over \bar{A}^2
\epsilon_M} \left[ 1 + {1 \over 4}\epsilon_M \right].}
Using \abar\ and \expans\ we see that these two surface terms are equal.
Thus,  taking the limit
$\epsilon_E \rightarrow 0$, the surface term in the Hamiltonian
vanishes. Since the volume term vanishes by virtue of the equations
of motion, this implies that the Hamiltonian vanishes for the Ernst
solution, and thus the boost energy is unchanged.

\newsec{Action and Area}

\subsec{The basic relations}

The fact that the boost energy is unchanged in the pair creation of
gravitating objects implies a simple relation between the Euclidean
action $I$ and the area of the horizons.  The Euclidean action is
 defined only with respect to a choice of background spacetime. If both
the background spacetime and original spacetime have acceleration horizons,
it is shown in  \haho\ that \impeq\ is modified to
\eqn\actiononext{ I =\beta H - {1 \over 4} (\Delta \A + \A_{bh}),}
where $\Delta \A$ is the difference in the area of the acceleration
horizon in the physical spacetime and the background.
Thus, for  the case of pair creation of
nonextremal black holes, we have
\eqn\actionnonext{ I_{Ernst} =\beta H_E - {1 \over 4} (\Delta \A + \A_{bh}),}
where $\Delta \A$ is the difference between the area of the acceleration
horizon in the Ernst metric and in the Melvin metric.
In the extreme case,  as shown in section 2, the area of the black hole horizon
does not appear in the action since the horizon is infinitely far away.
  Therefore the
action is given by
\eqn\actionext{ I_{Ernst} =\beta H_E - {1 \over 4} \Delta \A.}
We have shown that $H_E = 0$ in the previous section. The Ernst action is thus
\eqn\Hamreln{ I_{Ernst} = -{1 \over 4} (\Delta \A + \A_{bh})}
for the non-extreme case, and
\eqn\Hamrelnext{I_{Ernst} = -{1 \over 4}\Delta \A}
in the extreme case. We will now show that these relations in fact hold.

The area of the black hole event horizon in the Ernst solution can be
easily shown to be
\eqn\areabh{ {\cal A}_{bh} = \int_{y=\xi_4}
\sqrt{g_{xx}g_{\varphi\varphi}} dx d\varphi
= {\Delta \varphi_E (\xi_4-\xi_3) \over A^2 (\xi_3-\xi_2) (\xi_4 -
\xi_2)}}
where $\Delta \varphi_E$ is given in \phiperiod. We now turn to the calculation
of the other two terms in \Hamreln.

\subsec{Change in area of the acceleration horizon}

Since the acceleration horizon is non-compact, its area is infinite;
to calculate the difference, we must introduce a boundary, as we did
in calculating the Hamiltonian. If we introduce a boundary in the
Ernst solution at $x = \xi_3 + \epsilon_E$, the area of the region
inside it is
\eqn\areaE{ \eqalign{{\cal A}_E = &\int_{y=\xi_3} \sqrt{g_{xx}
g_{\varphi\varphi}}  dx d\varphi = {\Delta \varphi_E \over A^2}
\int_{x=\xi_3+\epsilon_E}^{x=\xi_4} {dx \over (x-\xi_3)^2} \cr &
= - {\Delta \varphi_E \over A^2 (\xi_4 - \xi_3)} +
{\Delta \varphi_E \over A^2 \epsilon_E} = - {4\pi L^2 \over A^2
G'(\xi_3) (\xi_4 - \xi_3)} + \pi \rho_E^2,}}
where we have used $L = \Lambda(\xi_3)$ and \phiperiod, and defined
$\rho_E^2 =  4L^2 /( A^2 G'(\xi_3) \epsilon_E)$. The
acceleration horizon in the Melvin solution is the surface $z=0$, $t=0$
in \melvins (this can be seen by introducing the Rindler-type coordinates
$t = \eta \sinh \hat t,\ z= \eta \cosh \hat t$). Its area inside a boundary at
$\rho=\rho_M$ may similarly be calculated to be
\eqn\areaM{ {\cal A}_M = \int \sqrt{g_{\rho\rho}
g_{\varphi\varphi}} d\rho d\varphi = 2\pi \int_{\rho=0}^{\rho=\rho_M}
\rho d \rho = \pi \rho_M^2.}
Note that there is no ambiguity in the choice of boundary in the
Melvin solution here; $\rho = \rho_M$ is the only choice which obeys
the Killing symmetry.

We must now match the intrinsic features of the boundary; we require
that the proper length of the boundary and the integral of the gauge
potential $A_\varphi$ around the boundary be the same. For the Ernst
solution, the proper length of the boundary is
\eqn\lengthE{ l_E = \int \sqrt{g_{\varphi\varphi}} d\varphi = {8 \pi
\over \widehat{B}_E^2 \rho_E} \left[ 1 - {4 \over \widehat{B}_E^2
\rho_E^2} - {L^2
G''(\xi_3) \over G'(\xi_3)^2 A^2} {1 \over \rho_E^2}\right].}
As in section 4, we expand to second non-trivial order in
$\rho_E$; higher-order terms do not affect $\Delta {\cal A}$ in the
limit $\rho_E \rightarrow \infty$. For the Melvin solution, the proper
length of the boundary is
\eqn\lengthM{ l_M = {8 \pi \over \widehat{B}_M^2 \rho_M} \left[ 1 - {4 \over
\widehat{B}_M^2 \rho_M^2} \right].}
The integral of the gauge potential around the boundary is, in the Ernst
solution,
\eqn\potE{ {1\over 2\pi} \oint A_\varphi d\varphi  =
   {2 \over \widehat{B}_E} - {2 A^2
\epsilon_E G'(\xi_3)
\over \widehat{B}_E^3 L^2}=  {2 \over \widehat{B}_E} - {8 \over
\widehat{B}_E^3 \rho_E^2},}
while in the Melvin solution it is
\eqn\potM{ {1\over 2\pi} \oint A_\varphi d\varphi =
  {2 \over \widehat{B}_M} - {8 \over
\widehat{B}_M^3  \rho_M^2}.}
If we write
\eqn\expansion{ \widehat{B}_M = \widehat{B}_E \left( 1+ {\beta \over
\rho_E^2} \right)
{\rm\ \ and\ \ }  \rho_M = \rho_E \left( 1+ {\alpha \over \rho_E^2 }
\right),}
then setting the integral of the gauge fields equal gives $\beta=0$,
as before, and setting $l_E=l_M$ perturbatively gives
\eqn\alphares{ \alpha = {L^2 G''(\xi_3) \over G'(\xi_3)^2 A^2}.}
Substituting this into \areaM\ gives
\eqn\aMafter{ {\cal A}_M =  \pi \rho_E^2 + 2 \pi \alpha = \pi \rho_E^2
+ {2 \pi L^2 G''(\xi_3) \over G'(\xi_3)^2 A^2}.}

We can now evaluate the difference in area, letting $\rho_E
\rightarrow \infty$,
\eqn\areadiff{ \eqalign{ \Delta {\cal A} = {\cal A}_E - {\cal A}_M &=
-{4 \pi L^2 \over G'(\xi_3)A^2}
\left[ {1\over (\xi_4-\xi_3)} + {G''(\xi_3) \over 2 G'(\xi_3)} \right]
\cr &
= - {4\pi L^2 \over G'(\xi_3) A^2} \left[ {1 \over (\xi_3-\xi_2)} + {1 \over
(\xi_3-\xi_1)} \right].}}
Now for the extreme case, $\xi_2=\xi_1$, and so
\eqn\areaext{ - {1 \over 4} \Delta {\cal A}
= {2 \pi L^2 \over G'(\xi_3) A^2 (\xi_3-\xi_1)},}
which agrees with the expression for the action found in \dggh. For the
non-extreme case,
\eqn\areanonext{ \eqalign{  - {1 \over 4} (\Delta {\cal A} +
{\cal A}_{bh})  & ={\pi L^2 \over G'(\xi_3) A^2} \left[ {2 \over
(\xi_3-\xi_1)} + {(\xi_2-\xi_1) \over (\xi_3-\xi_2) (\xi_3-\xi_1)} -
{(\xi_4-\xi_3) \over (\xi_4-\xi_2) (\xi_3- \xi_2) }\right] \cr & = {2
\pi L^2 \over G'(\xi_3) A^2(\xi_3 - \xi_1)},\cr}}
where we have used \areabh\ in the first step and the no-strut
condition \nostrtt\ in the second.
Notice that the final expression is the same as in \areaext.
So the relations \keyeq\ and \keyeqq\ are confirmed
provided the formula for the instanton action given in \dggh\ is valid for
both the extreme and nonextreme black holes. We now
verify that this is indeed the case.

\subsec{Direct calculation of the action}

In \dggh\ it was assumed that the divergent part of the action could
simply be subtracted, using a coordinate matching condition,  without
affecting the correct finite contribution to the action. As we have seen
above, this is not necessarily the case; we need to evaluate the action for
a bounded region, impose some geometric matching conditions at the
boundary to ensure that the boundaries are the same, and then let the
boundary tend to infinity.
Despite all this, the fact that our result above agrees with that in
\dggh\ suggests that the answer is unchanged, as we shall see.

To evaluate the action directly, we introduce a boundary 3-surface at
large radius.  We will take the surface to lie at $x-y= \epsilon_E$ in
the Ernst solution and at $x-y= \epsilon_M$ in the accelerated
coordinate system in the Melvin solution, as in section 4. The volume
integral of $R$ is zero by the field equations. The volume integral of
the Maxwell Lagrangian $F^2$ is not zero, but it can be converted to a
surface term and combined with the extrinsic curvature term,
as shown in \dggh. Thus the action of the region of the
Ernst solution inside the surface is made up of two parts:
boundary contributions from the 3-surface embedded in the Ernst
solution, and a subtracted contribution from the 3-surface embedded in
the Melvin solution.

The contribution to the action from this surface in the Ernst solution
is \dggh
\eqn\actionpE{ I_E = -{1 \over 8 \pi} \int_{x-y=\epsilon_E} d^3x
\sqrt{h} e^{-\delta} \nabla_\mu (e^\delta n^\mu) = {\pi L^2 \over A^2
G'(\xi_3)} \left[ -{3 \over \epsilon_E} + {2 \over (\xi_3 -\xi_1)}
\right],}
where $e^{-\delta} = \Lambda {(y-\xi_1) \over (x-\xi_1)}$, and $h$ is the
induced  metric on the 3-surface. The contribution from the surface in
the Melvin solution may be obtained by setting $r_+ = r_- =0$
in \actionpE; it is
\eqn\actionpM{ I_M = -{\pi \over 2 \bar{A}^2} {3 \over
\epsilon_M} + O(\epsilon_M).}

The matching conditions on the boundary follow immediately from the
conditions used to compute the Hamiltonian in section 4.
If we make the change of coordinates \cchangei\ and \intchii\
in the Ernst solution, and analytically continue $\tau' = it'$
then the induced metric on
the 3-surface in the Ernst solution is $^{(3)} ds^2 = N^2 d\tau'^2 +
{}^{(2)} ds^2$, where $N$ is the lapse \lapE\ and $^{(2)} ds^2$ is
given by \bmetricE. Similarly, if we use \chiM\ in
the Melvin solution \accmel\ and analytically continue $\tau = it$,
the induced metric on the 3-surface in the Melvin
solution will be $^{(3)} ds^2 = N^2 d\tau^2 + {}^{(2)} ds^2$, where $N$
is the lapse, given by \lapM, and $^{(2)} ds^2$ is given by
\bmetricM. The Maxwell field on
the 3-surface will be the same as in section 4. Therefore, we see that
the intrinsic features of the 3-surface may be matched by taking
\abar\ and \expans\ with $\alpha=\beta=0$.
The action may now be evaluated,
\eqn\diffaction{ I_{Ernst} = I_E - I_M = {\pi L^2 \over A^2 G'(\xi_3)}
\left[ -{3
\over \epsilon_E} - {3 G''(\xi_3) \over G'(\xi_3) \epsilon_M} + {2
\over (\xi_3 - \xi_1)} \right] = { 2 \pi L^2 \over A^2 G'(\xi_3)
(\xi_3-\xi_1)}.}
This applies to both extremal and nonextremal instantons and
agrees with the previous expressions in the literature.
One can understand why the naive coordinate subtraction of divergences \dggh\
yielded the correct answer since the  boundary geometry is
matched when $\alpha=0$.
Since \diffaction\ agrees
 with \areaext\ and \areanonext, we see that the
relations \keyeq\ and \keyeqq\ have been verified.

\newsec{Black Hole Annihilation}

As discussed in the introduction, since black holes can be pair created,
it must be possible for them to annihilate. This provides a new way
for black holes to disappear which does not involve Planck scale
curvature. The closest analog of the pair creation process is black hole
annihilation in the presence of a background magnetic field.
To reproduce the time reverse of pair creation
exactly one would have to arrange that the black holes had exactly the right
velocities to come to rest in a magnetic field at a critical separation. They
could then tunnel quantum mechanically and annihilate each other.
If the black holes came
to rest too far apart their total energy would be negative and they would not
be able to annihilate. If they were too near together it would still be
possible for them to annihilate but now there would be energy left over which
would be given off as electromagnetic or gravitational radiation.
It is also  possible for black holes to annihilate in the absence of a
magnetic field, with all of their energy converted to radiation.

One might ask whether the generalized second law of thermodynamics is
violated in this process. The answer is no. Even though the total
entropy is decreased by the elimination of the black hole horizons, this
is allowed since it is a rare process. The rate can be estimated as follows.
Nonextremal black holes behave in pair creation as if they had
$e^S$ internal states. Since two nonextremal black holes can presumably
annihilate only if they are in the same internal state, if one throws
two randomly chosen black holes together the probability
of direct annihilation is of order $e^{-S}$.

We argued in section 2 that extreme black holes are fundamentally different
from non-extreme holes since they have zero entropy. This presumably
implies that  two oppositely charged
extreme  black holes cannot form a neutral black hole. Instead, they
always directly annihilate. This is consistent with the idea that
extreme black holes cannot be formed in gravitational collapse, but can
only arise through pair creation.

The process of black hole annihilation also seems to violate the idea
that `black holes have no hair'. It would appear that one could determine
something about the internal state of a black hole, i.e., whether two black
holes are in the same state or different states, by bringing them together
and seeing if they annihilate. However, it is not clear how robust the
internal state is.
It is possible that simply the act
of bringing the black holes together will change their state.

The fact that the pair creation of nonextremal black holes creates a wormhole
in space could be taken as a  geometric manifestation of their correlated
state. However, we do not believe that black holes need to be connected
by wormholes in order to annihilate.  Imagine two pairs of
black holes being created. If each pair annihilates separately, the instanton
will contain two black hole loops, and one expects the action will be smaller
than that of extremal black holes (or gravitating monopoles) by twice the
black hole entropy. However, there should be another instanton in which
the two pairs are created and then the black holes from one pair
annihilate with those from the other. This instanton will contain one
black hole loop and will presumably have an action which is smaller by
one factor of the black hole entropy. This can be interpreted as arising
from  a contribution of minus twice the black hole entropy
from the  pair creation of the
two pairs, and a contribution of plus the black hole entropy from the
annihilation of one pair. (After one pair annihilates, the other pair
must be correlated, and does not contribute another factor of the black
hole entropy.)

It should be pointed out that even though the nonextremal black holes are
created with their horizons identified, it is still possible for them
to evolve independently. In particular, their horizon areas need not
remain equal. This is because the identification only requires that the
interior of the two black holes be the same. On a nonstatic slice which
crosses the future event horizon in Ernst, there are two separate horizons.
If one throws matter into one but not the other, the areas of the
two horizon components will not be equal at later times.
The fact that the horizon components share a common interior region of
spacetime
suggests
that the `internal' states of a black hole should be associated
with the region near the horizon.
Presumably, throwing matter into the holes will tend to decorrelate
 their `internal' states, but it is not clear whether just one particle is
enough to decorrelate them completely, or whether that requires a large number.

\vskip 1cm

\centerline{\bf Acknowledgements}

\vskip .3cm

This work was supported in part by NSF Grant PHY-9008502 and EPSRC Grant
GR/J82041. SFR acknowledges the financial support of the Association
of Commonwealth Universities and the Natural Sciences and Engineering
Research Council of Canada.

\appendix{A}{Generalization to include a dilaton }

\subsec{The dilaton Ernst solution}
The above investigations of the Ernst solution can be readily
extended to include a dilaton, as we now show. We consider the
general action
\eqn\actwa{ I = {1\over 16\pi} \int d^4 x [-R+ 2(\nabla \phi)^2
+e^{-2a\phi} F^2]  -{1\over 8\pi} \int K }
which has a parameter $a$ governing the strength of the dilaton
coupling. The Melvin and Ernst solutions are extrema of \actwa\ with $a=0$
and $\phi$ constant.
The generalization of the
Melvin solution to $a \neq 0$, first found by Gibbons and Maeda \gm,
is
\eqn\dmelv{
\eqalign{
&ds^2=\Lambda^{2\over 1+a^2}\left[-dt^2+dz^2+d\rho^2\right]
+\Lambda^{-{2\over 1+a^2}}\rho^2d\varphi^2,\cr
&e^{-2a\phi}=\Lambda^{2a^2\over 1+a^2},\qquad
A_\varphi={\widehat{B}_M\rho^2\over 2\Lambda},\cr
&\qquad \Lambda=1+{(1+a^2)\over 4}\widehat{B}_M^2\rho^2\ .\cr}
}
The generalization of the Ernst solution to this case is \dgkt\
\eqn\dernst{
\eqalign{
&ds^2=(x-y)^{-2}A^{-2}\Lambda^{2\over 1+a^2}
\left[F(x)\left\{G(y)dt^2-G^{-1}(y)dy^2\right\}
+F(y)G^{-1}(x)dx^2\right]\cr &\qquad +
(x-y)^{-2}A^{-2}\Lambda^{-{2\over 1+a^2}}F(y)G(x) d\varphi^2,\cr
&e^{-2a\phi}=e^{-2a\phi_0}\Lambda^{2a^2\over 1+a^2}
{F(y)\over F(x)},\quad
A_\varphi=-{2e^{a\phi_0}\over (1+a^2)B\Lambda}\[1+{(1+a^2)\over 2}Bqx\]
+k,\cr
}
}
where the functions $\Lambda \equiv\Lambda(x,y)$,
 $F(\xi)$ and $G(\xi)$ are now given by
\eqn\fns{\eqalign{
&\Lambda=\[1+{(1+a^2)\over 2}Bqx\]^2+{(1+a^2)B^2\over 4A^2(x-y)^2}
G(x)F(x), \cr
&F(\xi)=(1+r_-A\xi)^{2a^2\over (1+a^2)}\ ,\cr
&G(\xi)=(1-\xi^2-r_+A\xi^3)(1+r_-A\xi)^{(1-a^2)\over
(1+a^2)}\ ,\cr
}}
and $q^2 = r_+r_-/ (1+a^2)$. Here it is useful to define another
function,
\eqn\hdef{ H(\xi) \equiv G(\xi) F(\xi) = -(r_+A)(r_-A)
(\xi-\xi_1)(\xi-\xi_2)(\xi-\xi_3)(\xi-\xi_4),}
where $\xi_1 = -1 /( r_- A)$, and $\xi_2, \xi_3, \xi_4$ are the
roots of the cubic factor in $G(\xi)$. These roots have the same
interpretation as in the Ernst solution.

We now define $L = \Lambda ^{1 \over 1+a^2} (\xi_3)$, and set $k = 2
e^{a \phi_0} /B L^{1+a^2 \over 2}(1+a^2)$.  We then find that the
physical magnetic field and charge are \dggh
\eqn\bphys{\widehat{B}_E=
{B G^{\prime}(\xi_3)\over 2 L^{3+a^2 \over 2}}}
and
\eqn\pchge
{\widehat q = q { e^{a \phi_0} L^{3-a^2\over 2} (\xi_4-\xi_3) \over
G'(\xi_3) (1+ {1+a^2 \over 2} qB \xi_4)} \ .}

We restrict $x$ to the range $\xi_3 \leq x \leq \xi_4$ to get the
right signature. We have to impose the condition
\eqn\nonodesa{G^\prime(\xi_3)\Lambda(\xi_4)^{2\over 1+a^2} = -
G'(\xi_4) \Lambda(\xi_3)^{2 \over 1+a^2}}
to ensure that the conical singularities at both poles are eliminated
by choosing the period of $\varphi$ to be \phiperiod.
Setting the black hole parameters $r_+, r_-$ to zero in the dilaton Ernst
metric \dernst\ yields
\eqn\accdmel{\eqalign{ds^2 = &{\Lambda^{2 \over 1+a^2}
\over A^2 (x-y)^2} \left[ (1-y^2)
 dt^2 - {dy^2\over  (1-y^2)} + {dx^2\over
(1-x^2) }\right]\cr & + \Lambda^{-{2 \over 1+a^2}}{1-x^2 \over  (x-y)^2 A^2}
d\vp^2,}}
with
\eqn\Llimg{\Lambda = 1+ {(1+a^2)\widehat{B}_E^2\over 4} {1-x^2 \over A^2
 (x-y)^2}\ ,}
which is the dilaton Melvin solution \dmelv\ written in accelerated
coordinates.  The dilaton Ernst solution \dernst\  reduces to the
dilaton Melvin solution \dmelv\ at large spatial distances, $x , y
\rightarrow \xi_3$.

We obtain the Euclidean section  by setting $\tau = it$. In
the non-extremal case, $\xi_1 < \xi_2$,  we are forced to restrict
$\xi_2 \leq y \leq \xi_3$, and we find that to eliminate the conical
singularities, we have to choose the period of $\tau$ to be \pert\ and
impose the condition \nost, which gives
\eqn\roots{
\left({\xi_2-\xi_1\over \xi_3-\xi_1}\right)^{1-a^2\over 1+a^2}(\xi_4-\xi_2)
=(\xi_4 - \xi_3) }
for this metric. In the extremal case, the black hole horizon
$y=\xi_2$ is at an infinite distance, so the range of $y$ is $\xi_2 <
y \leq \xi_3$, and we only need to choose the period of $\tau$ to be
\pert\ to eliminate the conical singularity. The non-extremal
instanton still has topology $S^2
\times S^2 - \{pt\}$, while the extremal one has topology $R^2 \times
S^2 -\{pt \}$, and they have the same interpretation as before.

\subsec{Boost energy}

We now show that the boost energy is unchanged by pair creation in
this case. The Hamiltonian is still given by \hamil, and the volume
term vanishes, so it is just given by the surface term. We choose the
boundary in the Ernst solution to be given by $x-y = \epsilon_E$, and
make the coordinate transformations \cchangei\ and \intchii. In the
Melvin solution, we assume that the boundary has the form
\eqn\chiMg{x = -1 + \epsilon_M \chi(1 + \epsilon_E f(\chi)),\ \  y = -1
+\epsilon_M (\chi-1) (1+\epsilon_E g(\chi)),}
in the coordinates of the accelerated form \accdmel. In this case, we
need to match the value of the dilaton on the boundary, as well as the
induced metric and gauge field on the boundary. For the Ernst metric,
the induced metric on the boundary is
\eqn\bmetEg{\eqalign{^{(2)} ds^2 =& {2 L^2 F(\xi_3) \over A^2
\epsilon_E G'(\xi_3)}
\left\{ - {\lambda^{2 \over 1+a^2} d \chi^2 \over 2\chi (\chi-1)}
\left[ 1+ \epsilon_E (2 \chi -1) {F'(\xi_3) \over F(\xi_3)} \right]
\right. \cr & \left.  +
2 \lambda^{-{2 \over 1+a^2}} \chi \left[1 + \epsilon_E \chi {H''(\xi_3) \over
2 H'(\xi_3)} -\epsilon_E {F'(\xi_3) \over F(\xi_3)} \right] d
\varphi'^2 \right\}, }}
where
\eqn\lambdaEg{\lambda =  {(1+a^2)\widehat{B}_E^2 F(\xi_3)  L^2 \chi \over
A^2 G'(\xi_3) \epsilon_E} \[ 1 + \epsilon_E \chi
 { H''(\xi_3) \over 2 H'(\xi_3)} \] +1.}
The electromagnetic field on the boundary for the Ernst solution is
\eqn\fluxEg{F_{\chi\varphi'} = {2 L^2 F(\xi_3) \widehat{B}_E \over A^2
\epsilon_E G'(\xi_3) \lambda^2}{e^{a \phi_0} \over L^{a^2}}
 \left[ 1 + \epsilon_E \chi {H''(\xi_3) \over  H'(\xi_3)}  \right],}
and the dilaton at the boundary is
\eqn\dilbE{e^{-2a\phi} = e^{-2a \phi_0} L^{2a^2} \lambda^{2a^2 \over
1+a^2} \left( 1 - \epsilon_E {F'(\xi_3) \over F(\xi_3)} \right).}

In the Melvin solution, the induced metric on the boundary is
\eqn\bmetMg{\eqalign{^{(2)} ds^2 = & -{\Lambda^{2 \over 1+a^2} \over 2
\chi (\chi-1) \bar{A}^2 \epsilon_M} \left[ 1 - \epsilon_E (\chi-1)
f(\chi) + \epsilon_E \chi g(\chi) \right. \cr & \left. -2
\epsilon_E \chi (\chi-1) (f'(\chi) -g'(\chi)) -2 \epsilon_E (\chi
f(\chi) - (\chi-1) g(\chi)) \right] d\chi^2  \cr &   +
{2 \Lambda^{-{2 \over 1+a^2}} \chi \over  \bar{A}^2 \epsilon_M}\left[ 1
-{1 \over 2} \epsilon_M \chi
+ \epsilon_E f(\chi) -2 \epsilon_E (\chi f(\chi) - (\chi-1) g(\chi))
\right] d\varphi^2, }}
where
\eqn\lambdaMg{\Lambda = 1+ { (1+a^2) \widehat{B}_M^2 \chi \over 2 \bar{A}^2
\epsilon_M} \left[ 1 - {1 \over 2} \epsilon_M \chi + \epsilon_E
f(\chi) - 2 \epsilon_E (\chi f(\chi) - (\chi-1) g(\chi))\right] .}
The gauge field on the boundary in Melvin is
\eqn\fluxMg{\eqalign{F_{\chi \varphi} = & {\widehat{B}_M \over
\bar{A}^2 \epsilon_M
\Lambda^2} \left[ 1 - \epsilon_M  \chi + \epsilon_E (\chi f'(\chi) +
f(\chi)) -2 \epsilon_E (\chi f(\chi) - (\chi-1) g(\chi)) \right. \cr &
\left.  -2 \epsilon_E
\chi (f(\chi) + \chi f'(\chi) - g(\chi) - (\chi-1) g'(\chi)) \right],}}
and the dilaton at the boundary is
\eqn\dilbM{ e^{-2 a \phi} = \Lambda^{2 a^2 \over 1+a^2}.}

We fix the remaining coordinate freedom by taking
\eqn\abarg{\bar{A}^2 = -{G'(\xi_3) \over 2 L^2 F(\xi_3)}{H'(\xi_3)
\over H''(\xi_3)}
A^2,}
and write
\eqn\expang{
 e^{a\phi_0} = L^{a^2} \left(1 - \gamma  \epsilon_E
\right), \ \ \widehat{B}_M = \widehat{B}_E \left(1+\beta \epsilon_E
\right).}
We then find that the intrinsic metric, gauge field and dilaton on the boundary
can all be matched by taking
\eqn\subl{ \epsilon_M = - {H''(\xi_3) \over H'(\xi_3)} \epsilon_E,\ \
f(\chi) = {F'(\xi_3) \over F(\xi_3)} (4\chi -3),\ \ g(\chi) = {F'(\xi_3)
\over F(\xi_3)} (4\chi-1),}
and
\eqn\consts{\beta= \gamma = {1 \over 2} {F'(\xi_3) \over F(\xi_3)}.}

We should note that the lapse function is also matched by these
conditions. For the Ernst metric \dernst, the lapse function  is
\eqn\lapEg{N = \left(4 L^2 F(\xi_3) \over A^2 \epsilon_E
G'(\xi_3)\right)^{1 \over 2} \lambda^{1 \over 1+a^2} \sqrt{1- \chi} \left[ 1
+ {1 \over 4} \epsilon_E (\chi-1) {H''(\xi_3) \over H'(\xi_3)} + {1
\over 2} \epsilon_E {F'(\xi_3) \over F(\xi_3)} \right],}
While the lapse function for the Melvin metric \accdmel\ is
\eqn\lapMg{N = \left(2 \over \bar{A}^2
\epsilon_M\right)^{1 \over
2} \Lambda^{1 \over 1+a^2} \sqrt{1-\chi} \left[ 1 - {1 \over 4}
\epsilon_M (\chi-1) + {1 \over 2} \epsilon_E g(\chi) - \epsilon_E (
\chi f(\chi) - (\chi-1) g(\chi)) \right].}
We see that \subl\ and \consts\ make \lapEg\ and \lapMg\ equal as well.

The extrinsic curvature of the boundary embedded in the Ernst solution
is
\eqn\excurvg{^2 K = {A \epsilon_E^{1/2} G'(\xi_3)^{1/2} \over L
F(\xi_3)^{1/2} \lambda^{1 \over 1+a^2}} \left[ 1+ {1 \over 4}
\epsilon_E {H''(\xi_3) \over H'(\xi_3)} (4 \chi -3) - {1 \over
2}\epsilon_E {F'(\xi_3) \over F(\xi_3)} (4 \chi -3) \right],}
while the extrinsic curvature of the boundary embedded in the Melvin
solution is
\eqn\excurvMg{^2 K_0 = {\bar{A} \epsilon_M^{1/2} \sqrt{2} \over
\Lambda^{1 \over 1+a^2}} \left[ 1 -{1 \over 4} \epsilon_M (4\chi-3) -
{1 \over 2} \epsilon_E {F'(\xi_3) \over F(\xi_3)} (24\chi -13) \right].}
Using the matching conditions \abarg\ and \subl, we may now evaluate
\eqn\ecdiff{ ^2 K - {}^2 K_0 = {5 A \epsilon_E^{3/2} G'(\xi_3)^{1/2} \over L
F(\xi_3)^{1/2} \lambda^{1 \over 1+a^2}} {F'(\xi_3) \over F(\xi_3)}
(2\chi -1).}
Therefore, taking the limit $\epsilon_E \rightarrow 0$,  the Hamiltonian is
\eqn\hameval{ H_E  = -{1\over 4} \int_0^1 d \chi N \sqrt{h} (^2 K - {}^2
K_0) = - {5 L^2 F'(\xi_3) \over A^2 G'(\xi_3)} \int_0^1
d\chi (2 \chi -1) =0,}
where $h$ is the determinant of the metric (\bmetEg\ or
\bmetMg). Thus, \keyeq\ and \keyeqq\ still hold, which we will now
confirm by direct calculation.

\subsec{Horizon area and instanton action}

We begin by calculating the difference in area. The area of the black
hole is now given by
\eqn\areabho{ { \cal A}_{bh}= {F(\xi_2) \Delta \varphi_E (\xi_4-\xi_3) \over
A^2 (\xi_3-\xi_2) (\xi_4 - \xi_2)} = {4 \pi F(\xi_2) L^2 \over A^2
G'(\xi_3)} {(\xi_4-\xi_3) \over (\xi_3-\xi_2) (\xi_4-\xi_2)},}
and the area of the acceleration horizon in the Ernst solution, inside
a boundary at $x = \xi_3 + \epsilon_E$, is
\eqn\areaEo{{\cal A}_E = {F(\xi_3) \Delta \varphi_E \over A^2}
\int_{x=\xi_3+\epsilon_E}^{x=\xi_4}
{dx \over (x-\xi_3)^2} = - {4 \pi F(\xi_3) L^2  \over A^2 G'(\xi_3)
(\xi_4-\xi_3)} + \pi \rho_E^2, }
where $\rho_E^2 = {4 F(\xi_3) L^2 \over G'(\xi_3) A^2 \epsilon_E}$
now. Also,  ${\cal A}_M$ is still given by \areaM. The boundary
conditions in this case are that the proper length of the boundary,
the integral of the gauge potential around the boundary and the value
of the dilaton at the boundary are the same.

The proper length of the boundary in the dilaton Ernst solution is
\eqn\lengthEo{ \eqalign{l_E & = {4 \pi 2^{1-a^2 \over 1+a^2}
\rho_E^{a^2-1 \over
1+a^2} \over [(1+a^2)\widehat{B}_E^2]^{1 \over 1+a^2}} \left\{ 1 + {F(\xi_3)
L^2
\over G'(\xi_3) A^2} {1 \over \rho_E^2} \left[ {a^2-1 \over 1+a^2}
{H''(\xi_3) \over H'(\xi_3)} - {2 F'(\xi_3) \over F(\xi_3)} \right]
\right. \cr & \left. - {4
\over \widehat{B}_E^2 \rho_E^2 (1+a^2)^2} \right\},}}
and the proper length of the boundary in the dilaton Melvin solution is
\eqn\lengthMo{ l_M = { 4 \pi 2^{1-a^2 \over 1+a^2} \rho_M^{a^2-1 \over
1+a^2} \over [(1+a^2)\widehat{B}_M^2]^{1 \over 1+a^2}} \left[ 1 - {4 \over
\widehat{B}_M^2 \rho_M^2 (1+a^2)^2} \right].}
It is interesting to note that the proper length behaves quite
differently for $a^2<1$ and $a^2>1$. The integral of the gauge potential
around the boundary curve is, in the dilaton Ernst solution
\eqn\potEo{ \oint A_\varphi d \varphi = {4 \pi e^{a \phi_0} \over
(1+a^2)  L^{a^2} \widehat{B}_E} \left[ 1 - {4 \over \widehat{B}_E^2 \rho_E^2
(1+a^2)} \right],}
while for the dilaton Melvin solution it is
\eqn\potMo{ \oint A_\varphi d \varphi = {4 \pi \over (1+a^2)
\widehat{B}_M} \left[ 1 - {4
\over \widehat{B}_M^2 \rho_M^2 (1+a^2)} \right].}
Finally, the dilaton field at the boundary is, for the dilaton Ernst
solution
\eqn\scalarE{\eqalign{ e^{-2a\phi} &= e^{-2a \phi_0} \Lambda^{2a^2 \over 1+a^2}
{ F(\xi_3) \over F(\xi_3 + \epsilon_E)} \cr &= e^{-2a \phi_0} L^{2 a^2} \left[
(1+a^2) \widehat{B}_E^2 \rho_E^2 \over 4 \right]^{2a^2 \over 1+a^2} \left[ 1 +
{4a^2 \over 1+a^2 } {F(\xi_3) L^2 H''(\xi_3) \over G'(\xi_3) A^2
H'(\xi_3)} {1 \over \rho_E^2} \right. \cr & \left.
+ {8a^2 \over (1+a^2)^2} {1 \over \widehat{B}_E^2
\rho_E^2} - {4 F'(\xi_3) L^2 \over G'(\xi_3) A^2} {1 \over \rho_E^2} \right],}}
while for dilaton Melvin it is
\eqn\scalarM{ e^{-2a \phi} = \left[ (1+a^2) \widehat{B}_M^2 \rho_M^2 \over 4
\right]^{2a^2 \over 1+a^2} \left[ 1 + {8 a^2 \over (1+a^2)^2} {1 \over
\widehat{B}_M^2 \rho_M^2} \right].}

We now write
\eqn\expano{ \widehat{B}_M = \widehat{B}_E \left(1+{\beta \over \rho_E^2}
\right), \ \ \rho_M
= \rho_E \left( 1+{\alpha \over \rho_E^2} \right),}
and
\eqn\expanoo{ e^{a\phi_0} = L^{a^2} \left(1 - {\gamma \over \rho_E^2}
\right).}
We may solve for $\alpha, \beta$ and $\gamma$ by setting the
various quantities equal perturbatively. This gives
\eqn\alphao{ \alpha = {F(\xi_3) L^2 \over G'(\xi_3) A^2} \left[
{H''(\xi_3) \over H'(\xi_3)} - {2 F'(\xi_3) \over F(\xi_3)} \right],}
$$ \beta= \gamma = {2 F'(\xi_3) L^2 \over G'(\xi_3) A^2}. $$

We may now calculate the difference in area:
\eqn\areadiffo{\eqalign{ \Delta {\cal A} &= -{4\pi L^2 F(\xi_3)
\over A^2  G'(\xi_3) (\xi_4 - \xi_3)} -2\pi \alpha \cr &
= -{4\pi L^2 F(\xi_3) \over A^2 G'(\xi_3)} \left[ {1 \over
(\xi_4-\xi_3)} + { H''(\xi_3) \over 2 H'(\xi_3)} - {F'(\xi_3) \over
F(\xi_3)} \right] \cr &
= -{4 \pi L^2 F(\xi_3) \over A^2 G'(\xi_3)} \left[ {(\xi_2-\xi_1) \over
(\xi_3-\xi_1) (\xi_3 - \xi_2)} + {F'(\xi_3) \over a^2 F(\xi_3)}
\right].}}
For the extreme case, $\xi_2 = \xi_1$, and so,
\eqn\areaexto{  - {1 \over 4} \Delta {\cal A} =
{\pi L^2 F'(\xi_3) \over a^2 A^2 G'(\xi_3)},}
which agrees with the expression for the instanton action in \dggh.
For the non-extreme case,
\eqn\areanonexto{\eqalign{ - {1 \over 4} (\Delta {\cal A} +
{\cal A}_{bh}) &
= {\pi L^2 \over A^2 G'(\xi_3)} \left[ {F'(\xi_3)
\over a^2} + { F(\xi_3)(\xi_2-\xi_1) \over (\xi_3-\xi_2) (\xi_3-\xi_1)} -
{F(\xi_2) (\xi_4-\xi_3) \over (\xi_4-\xi_2) (\xi_3- \xi_2) }\right]
\cr & = {\pi L^2 F'(\xi_3) \over a^2 A^2 G'(\xi_3)},}}
where we have used \roots\ to cancel the last two terms.

Now we turn to the direct calculation of the action. The contribution
to the action from a boundary at $x-y = \epsilon_E$ embedded in the
Ernst solution is \dggh
\eqn\actionEg{I_E = -{1 \over 8 \pi} \int_{x-y=\epsilon_E} d^3 x
\sqrt{h} e^{-\phi/a} \nabla_{\mu} (e^{\phi/a} n^\mu) = {\pi L^2
 \over A^2 G'(\xi_3)} \left[ -{3 F(\xi_3) \over \epsilon_E} +
{F'(\xi_3) \over a^2} \right].}
As the solution is independent of $\tau$, the metric on this boundary is
just $^{(3)} ds^2 = {}^{(2)} ds^2 + N^2 d\tau^2$, where $^{(2)} ds^2$ is
given by \bmetEg, and the gauge field and dilaton on the boundary are
\fluxEg\ and \dilbE. Thus, if we assume the boundary in the Melvin
solution has the form \chiMg, then we may see that \subl\  and \consts\
will match the metric, gauge field and dilaton on the boundary. The
contribution to the action from the boundary embedded in the Melvin
solution is then
\eqn\aactionMg{\eqalign{I_M  &= -{1 \over 8 \pi} \int_{bdry.} d^3 x \sqrt{h}
e^{-\phi/a} \nabla_\mu ( e^{\phi/a} n^\mu) \cr & = {\pi \over 8 \bar{A}^2
\epsilon_M} \int_0^1 d \chi \left[ -12 + 5 \epsilon_M (2\chi-1) + {103
F'(\xi_3) \over 2 F(\xi_3)} \epsilon_E (2\chi -1) \right] \cr & = -{3 \pi
\over 2 \bar{A}^2 \epsilon_M}.}}
Thus, using \abarg, we may evaluate the action,
\eqn\actiong{ I_{Ernst} = I_E - I_M = {\pi L^2 F'(\xi_3) \over A^2
G'(\xi_3) a^2},}
which is in perfect agreement with \dggh. As \actiong\ agrees with
\areaexto\ and \areanonexto, we have explicitly shown that \keyeq\ and
\keyeqq\ hold for general $a$.

\vfill\eject

\listrefs

\end